\documentclass{natureb}
\usepackage{graphicx}

\title{\textnormal{\textbf{Fingering versus stability in the limit of zero interfacial tension}}}

\author{Irmgard Bischofberger$^1$$^*$, Radha Ramachandran$^1$ \& Sidney R. Nagel$^1$}

\begin{document}

\maketitle

\begin{affiliations}
 \item The James Franck and Enrico Fermi Institutes and The Department of Physics, The University of Chicago, Chicago, Illinois 60637, USA
\end{affiliations}

\begin{abstract} 
The invasion of one fluid into another of higher viscosity in a quasi-two-dimensional geometry typically produces complex fingering patterns. Because interfacial tension suppresses short-wavelength fluctuations, its elimination by using pairs of miscible fluids would suggest an instability producing highly ramified singular structures. Previous studies focused on wavelength selection at the instability onset and overlooked the striking features appearing more globally. Here we investigate the non-linear growth that occurs after the instability has been fully established. We find a rich variety of patterns that are characterized by the viscosity ratio between the inner and the outer fluid, $\eta_{\mathrm{in}}$/$\eta_{\mathrm{out}}$, as distinct from the most-unstable wavelength, which determines the onset of the instability. As $\eta_{\mathrm{in}}$/$\eta_{\mathrm{out}}$ increases, a regime dominated by long highly-branched fractal fingers gives way to one dominated by blunt stable structures characteristic of proportionate growth. Simultaneously, a central region of complete outer-fluid displacement grows until it encompasses the entire pattern at $\eta_{\mathrm{in}}$/$\eta_{\mathrm{out}}\approx0.3$.
\end{abstract}

Viscous-fingering instabilities, created when a fluid confined between two parallel plates invades another of higher viscosity, have served for more than half a century as a prototype for complex pattern formation\cite{Saffman1958}. Such unstable interfaces are important in applications, significantly for sugar refining, oil recovery, hydrology and carbon sequestration\cite{Hill1952,Orr1984,Cinar2009,Gorell1983,Homsy1987} and much recent work has focused on how to control the instabilities\cite{Pihler2012,Lindner2000,Dias2013,Al2012}. Viscous fingering plays a central role in our understanding of pattern formation in part because it is amenable to both theory and experiment\cite{Saffman1958,Paterson1981,Bensimon1986,Homsy1987, Chen1989, Setu2013}. Of particular importance is the limit where the characteristic finger width, set by the most-unstable wavelength, $\lambda_{\mathrm{c}}$, approaches zero. In that case, it was expected that highly ramified fractal structures, similar to diffusion-limited aggregation, would form\cite{Sander1985,Witten1981,Paterson1984,Bettelheim2005}.

The most-unstable wavelength derived by Saffman and Taylor \cite{Saffman1958} is governed by a competition between interfacial tension, $\sigma$, and stresses that depend on the interfacial velocity, $V$, and the viscosity difference between the outer and inner fluids, $\Delta$$\eta$ $\equiv$ $\eta_{\mathrm{out}}$ - $\eta_{\mathrm{in}}$:
\begin{equation}
\lambda_{c}\ = \pi b \sqrt\frac{\sigma}{\Delta\eta V}
\label{eq:Cross}
\end{equation}
where $b$ is the gap thickness between the two parallel plates in the Hele-Shaw cell\cite{Saffman1958}. In this analysis, an instability occurs whenever $\Delta$$\eta$$>$$0$.

Lajeunesse \emph{et al.} studied the small wavelength limit in experiments using miscible fluids with negligible interfacial tension. They reported a complete suppression of the instability for fluids with $\Delta$$\eta$$>$$0$ when the viscosity ratio $\eta_{\mathrm{in}}$/$\eta_{\mathrm{out}}$ was above a certain threshold value\cite{Lajeunesse1997, Lajeunesse1999}. This result is highly counter intuitive --- it implies that the interface becomes stabilized when interfacial tension, which is expected to stabilize the interface, is removed.  
They further demonstrated that the onset of the instability at lower viscosity ratio coincides with a change in the profile of the interface across the gap: in the range of viscosity ratios where the interface remained stable the profile is quadratic near its tip, whereas in the unstable regime it is characterized by a flat shock front\cite{Lajeunesse1997, Lajeunesse1999}. 

To account for these experimental observations, Lajeunesse \emph{et al.} used the kinematic-wave-theory approach\cite{Yang1997} to describe analytically the interface profile across the gap. The theory successfully predicts the transition between a rounded and a shock profile with decreasing viscosity ratio, with the transition occurring at $\eta_{\mathrm{in}}$/$\eta_{\mathrm{out}}$ $=$ 0.67. While this analysis predicts the observed profiles across the gap, it does not explain the transition from stable to unstable displacement in the lateral direction that appears to be directly related to the formation of the shock front; why the shock front is seemingly indispensable for the instability to occur is an open question. In particular, it remains to be understood why the instability is suppressed in the range 0.67 $<$ $\eta_{\mathrm{in}}$/$\eta_{\mathrm{out}}$ $<$ 1, where the classical Saffman-Taylor analysis predicts the interface to be unstable. 

Both the prediction for the most-unstable wavelength and Lajeunesse's study on the threshold for the instability in miscible fluids concern the onset of the instability. Here, we focus instead on the non-linear regime characterizing the growth of the instability, and show that a large variety of patterns emerges. Our studies reveal a previously overlooked control parameter, the ratio of the viscosities of the inner and the outer fluid, that governs the large-scale structure of the patterns. 
In particular, we find at all values of $\eta_{\mathrm{in}}$/$\eta_{\mathrm{out}}$ that there is a central region of complete displacement of the outer fluid.  This region, which is very small at small $\eta_{\mathrm{in}}$/$\eta_{\mathrm{out}}$, grows larger with increasing viscosity ratio until it comprises the entire pattern. Thus, we show that the two previously reported regimes, the fractal growth regime of Saffman and Taylor and the regime of complete stability of Lajeunesse, occur at the two extremes of the viscosity ratio spectrum at low and high viscosity ratio.
Close to the boundary to the stable regime, we identify a novel type of behavior; here, the interface is initially stable before becoming unstable to small structures that, once formed, remain stable even though they grow much broader than the most unstable wavelength, $\lambda_{\mathrm{c}}$. We show that these structures exhibit features of a type of growth behavior very rarely observed in physical systems, known as proportionate growth.

\textbf{Results} \\
\textbf{Viscosity ratio sets large-scale features of instability.} 
For our experiments we use a radial Hele-Shaw cell of diameter $L$ = 28 cm with a typical gap spacing of $b$ = 254 $\mu$m. The fluids are injected through a hole in the center of one of the plates at a precise flow rate set by a syringe pump. We investigate the growth of patterns formed in the small-wavelength limit using pairs of miscible fluids, where the interfacial tension $\sigma$ is negligible. For large enough Peclet numbers, where advection dominates over diffusion, the inter-diffusion of the fluids is negligible so that the fluids remain separated by a well-defined interface\cite{Lajeunesse1999,Yang1997,DErrico2004}. This is the situation we investigate here.
Previous studies have shown that the most-unstable wavelength of Eq. 1 is cut off  at $\lambda_{\mathrm{c}}$ $\approx 5b$\cite{Paterson1985,Lajeunesse1997}, that is it is determined by the smallest length scale of the system, the plate spacing $b$.  Despite $\lambda_{\mathrm{c}}$ being identical for all our experiments, we find a variety of distinct large-scale structures, as shown in the series of images in Fig. 1a. Remarkably, these patterns are set by the \emph{ratio} of the two viscosities, in striking contrast to the most-unstable wavelength.

At low $\eta_{\mathrm{in}}$/$\eta_{\mathrm{out}}$, branched fractals are observed (panel 1 in Fig. 1a). With increasing viscosity ratio, however, an inner circle devoid of fingers comprising only the invading fluid appears and systematically grows larger (panel 2-4 in Fig. 1a). At high $\eta_{\mathrm{in}}$/$\eta_{\mathrm{out}}$, this inner circle comprises the entire pattern; the instability is fully suppressed (panel 5 in Fig. 1a).

\begin{figure}
\centering
\includegraphics[scale=0.5]{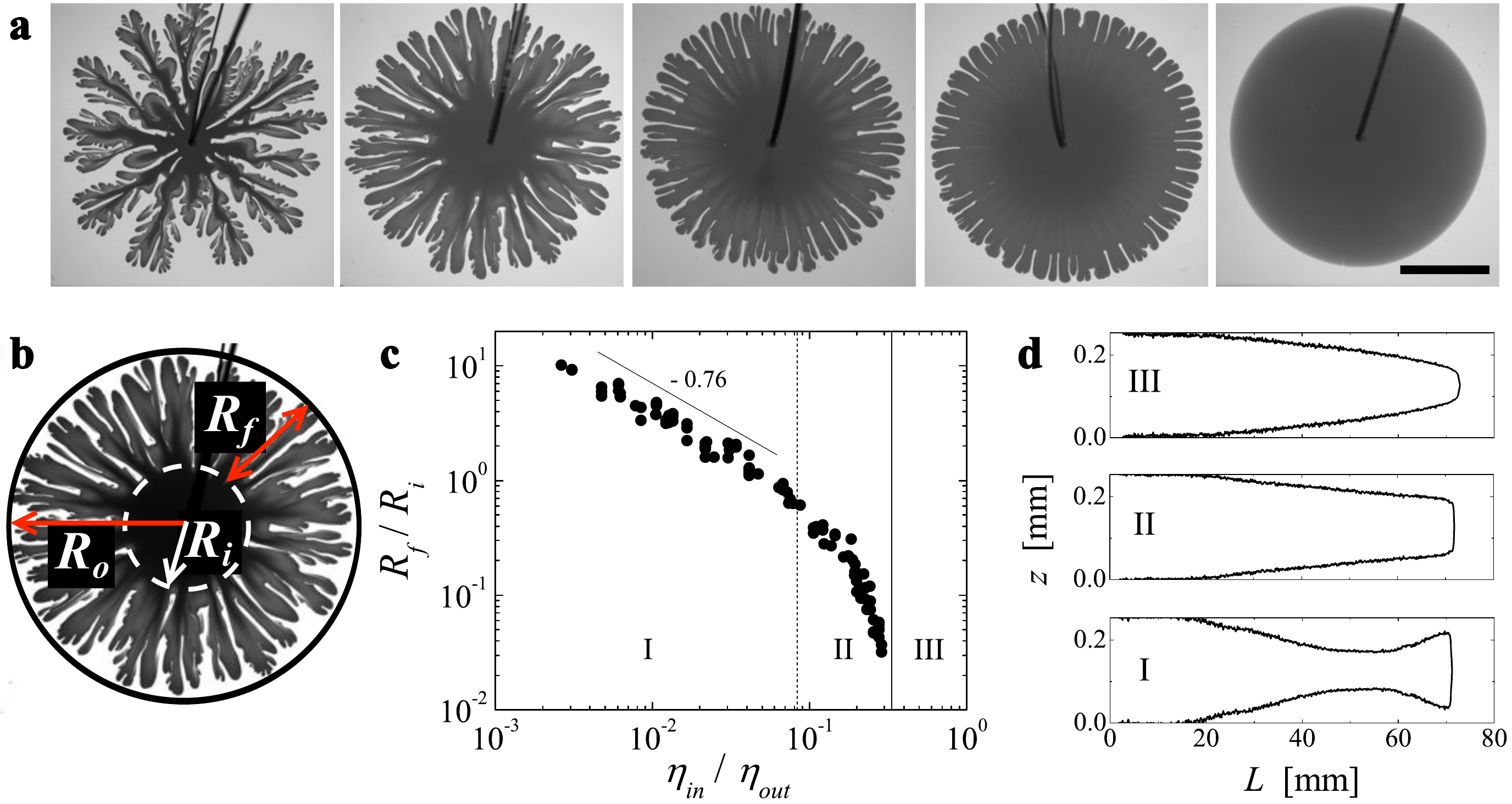}
\caption{ \textnormal{\textbf{Viscosity ratio as control parameter for miscible fingering patterns.} \textbf{a}, Fingering patterns for different viscosity ratios (from left to right): $\eta_{\mathrm{in}}$/$\eta_{\mathrm{out}}$ = 3.07$\cdot$10$^{-3}$, 3.10$\cdot$10$^{-2}$, 6.34$\cdot$10$^{-2}$, 1.23$\cdot$10$^{-1}$, 3.70$\cdot$10$^{-1}$.  As $\eta_{\mathrm{in}}$/$\eta_{\mathrm{out}}$ increases, the patterns become progressively less branched.  The scale bar is 5 cm. The flow rate is $q$ = 1 ml min$^{-1}$ and the gap $b$ = 254 $\mu$m.  \textbf{b}, Definition of the three characteristic length scales $R_{\mathrm{i}}$, $R_{\mathrm{o}}$ and $R_{\mathrm{f}}$.  \textbf{c}, The size ratio, $R_{\mathrm{f}}$/$R_{\mathrm{i}}$, versus $\eta_{\mathrm{in}}$/$\eta_{\mathrm{out}}$ when $R_{\mathrm{o}}$ = 80 mm.  Three regimes are identified: a fingering regime characterized by $R_{\mathrm{f}}$/$R_{\mathrm{i}}$ $\propto$ ($\eta_{\mathrm{in}}$/$\eta_{\mathrm{out}}$)$^{-0.76\pm0.05}$ (I), a transition regime where the instability occurs after a delay (II) and a stable regime (III).  The solid and dashed lines indicate the phase boundaries. \textbf{d}, For miscible fluids, the displacement of the outer fluid by the invading one produces three-dimensional tongues in the $z$-direction across the gap. Cross-section along the length $L$ of a tongue in each regime. From top to bottom, $\eta_{\mathrm{in}}$/$\eta_{\mathrm{out}}$ = 5.26$\cdot$10$^{-1}$, 2.04$\cdot$10$^{-1}$ and 1.11$\cdot$10$^{-2}$. The transition of the tongues from round to flat coincides with the instability onset.}}
\end{figure}

To quantify the structures, we define three characteristic lengths shown in Fig. 1b: an outer radius, $R_{\mathrm{o}}$, which is the radius of the smallest circle that encloses the entire fingering pattern, an inner radius, $R_{\mathrm{i}}$, which is the radius of the largest circle that fits inside the region where the outer fluid is fully displaced, and the finger length, $R_{\mathrm{f}}$ $\equiv$ $R_{\mathrm{o}}$ - $R_{\mathrm{i}}$.  Figure 1c shows three regimes when the ratio of the finger length to the inner radius, $R_{\mathrm{f}}$/$R_{\mathrm{i}}$, is plotted versus the viscosity ratio, $\eta_{\mathrm{in}}$/$\eta_{\mathrm{out}}$.  Regime I: For $\eta_{\mathrm{in}}$/$\eta_{\mathrm{out}}$ $<$ 0.083, the fingers form a fractal pattern; they elongate and split repeatedly creating new generations of fingers. 
Here we find $R_{\mathrm{f}}$/$R_{\mathrm{i}}$ $\propto$ ($\eta_{\mathrm{in}}$/$\eta_{\mathrm{out}}$)$^{-0.76\pm0.05}$. Regime II: For 0.083 $<$  $\eta_{\mathrm{in}}$/$\eta_{\mathrm{out}}$ $<$ 0.33, after an initial delay period small blunt structures form, which do not subsequently split.  
In this regime, $R_{\mathrm{f}}$/$R_{\mathrm{i}}$ decreases rapidly with $\eta_{\mathrm{in}}$/$\eta_{\mathrm{out}}$ and approaches zero at the boundary with the stable regime. Regime III: For $\eta_{\mathrm{in}}$/$\eta_{\mathrm{out}}$ $>$ 0.33, the interface between the two liquids is stable on the time- and length scale of the experiments.

The viscosity ratio further determines the interface profile across the gap in the $z$-direction.  The removal of interfacial tension allows for three-dimensional structures, ``tongues'' of one fluid penetrating into the other\cite{Lajeunesse1997,Lajeunesse1999,Yang1997,Goyal2006}. In regime III, the tip of the tongue is rounded as shown in the upper panel in Fig. 1d.  In contrast, the bottom two panels show that at lower viscosity ratios, the front of the tongue is flat and propagates as a shock front.  These profiles are characteristic of patterns formed in regimes II and I.  The shapes of the tongues agree qualitatively with the measurements of Lajeunesse \emph{et al.}\cite{Lajeunesse1997,Lajeunesse1999} for the miscible displacement in a linear Hele-Shaw cell in the presence of gravity. This transition in the profile of the interface at the tongue tip is clearly the most striking characteristic of the structure in the $z$-direction and corresponds to the position from which $R_{\mathrm{f}}$ and $R_{\mathrm{i}}$ are determined. Other finer details are present in the structure, as evidenced by the variety of intensity shadings within the inner fluid. However, we have not found a qualitative and dramatic change in this structure that delineates the three regimes we here identified.

\textbf{Delayed onset of instability in toe regime.} 
To explore the transition to the stable regime at high viscosity ratios, we investigate the temporal evolution in regime II.  Here, the onset of the instability is delayed: the interface is initially stable and only later develops small blunt structures, which we call ``toes''. The radial geometry is known to produce a small delay in the onset of the instability that also depends on the viscosity ratio between the two fluids\cite{Paterson1981,Nagel2013}.  However, the  delay calculated from this geometrical effect leads to an onset radius of only about 1 mm (comparable to the central injection hole in our plates) and is thus clearly not responsible for the much larger delay in the toe formation we observe. 
In this toe regime, after a delayed onset $R_{\mathrm{f}}$/$R_{\mathrm{i}}$ first increases rapidly and then slows down.  The inset of Fig. 2a shows that both the crossover between rapid and slow growth and the magnitude of $R_{\mathrm{f}}$/$R_{\mathrm{i}}$ depend on $\eta_{\mathrm{in}}$/$\eta_{\mathrm{out}}$. However, the main panel in Fig. 2a shows that all data in regime II can be scaled onto a single curve.  The collapse is obtained by normalising the time with a characteristic time, $t_{\mathrm{c}}$$^{*}$, and the size ratio with a characteristic size ratio, ($R_{\mathrm{f}}$/$R_{\mathrm{i}}$)$_{\mathrm{c}}$.   In these experiments, we also have varied the flow rate, $q$. We can normalise out the overall effect of the flow rate by defining: $t_{\mathrm{c}}$ $\equiv$ $t_{c}$$^{*}$/($q_{1ml/min}$/$q$) to account for the different flow rates used. Upon approaching the boundary to stable displacement, the instability onset becomes progressively delayed and the finger growth decreases in magnitude, as shown in Fig. 2b and c. Both $t_{\mathrm{c}}$ and ($R_{\mathrm{f}}$/$R_{\mathrm{i}}$)$_{\mathrm{c}}$ are independent of $q$. This confirms that effects due to diffusion of one liquid into the other are indeed irrelevant for the observed phenomena.

\begin{figure}
\centering
\includegraphics[scale=0.5]{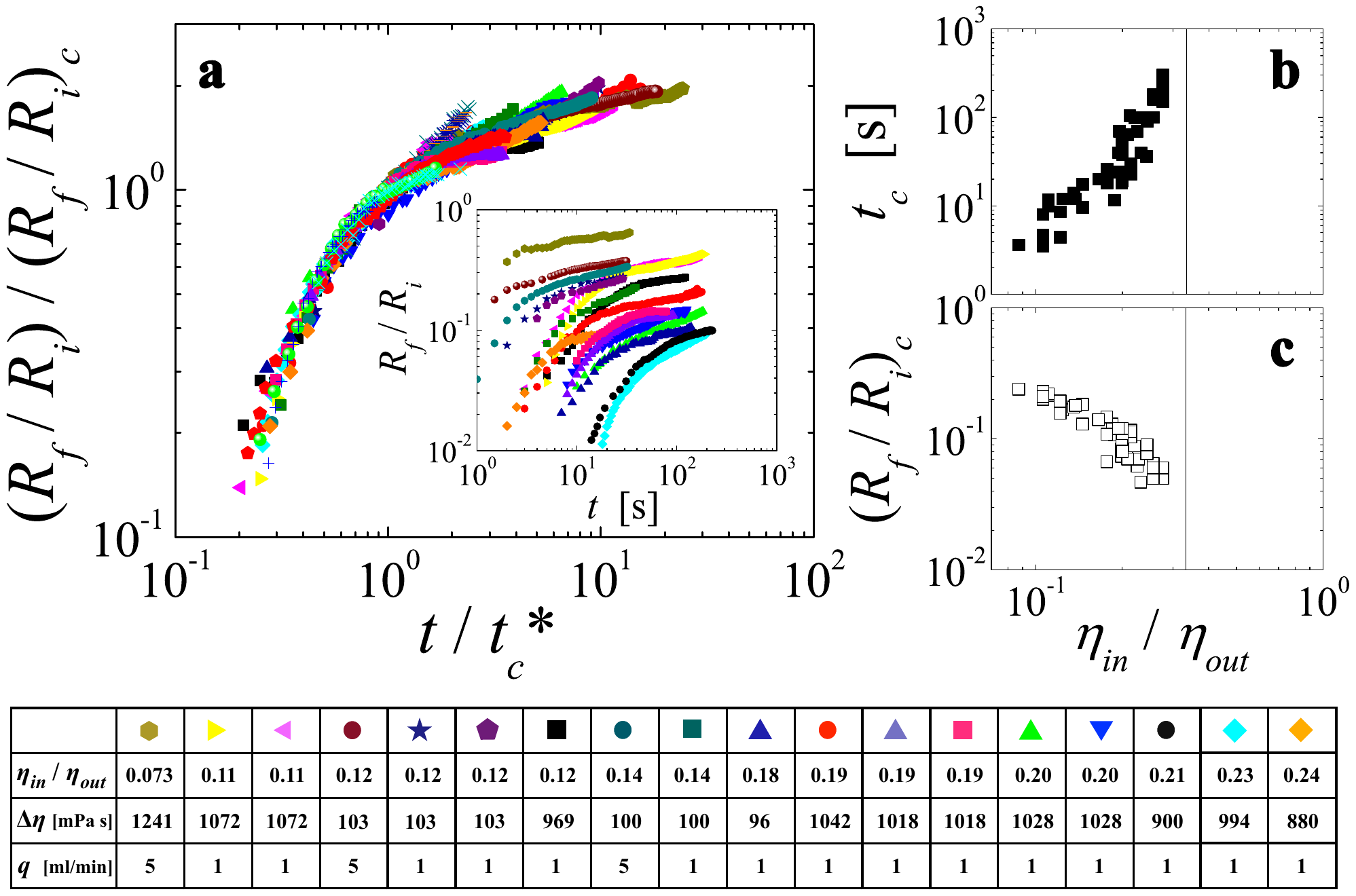}
\caption{ \textnormal{\textbf{Temporal evolution of the instability onset in regime II.} \textbf{a},  Scaled master curve of the size ratio $R_{\mathrm{f}}$/$R_{\mathrm{i}}$ versus time for patterns formed in regime II.  Pattern evolution is characterized by rapid growth at early times followed by very slow growth. $t_{\mathrm{c}}$$^{*}$ is defined as $t_{\mathrm{c}}$$^{*}$ = $t_{\mathrm{c}}$($q_{1ml/min}$/$q$)  Inset: $R_{\mathrm{f}}$/$R_{\mathrm{i}}$ versus time for experiments with different viscosity ratios and different flow rates. Symbols corresponding to the data sets are given in the table. \textbf{b}, Characteristic time $t_{\mathrm{c}}$ and \textbf{c}, characteristic size ratio ($R_{\mathrm{f}}$/$R_{\mathrm{i}}$)$_{\mathrm{c}}$ versus $\eta_{\mathrm{in}}$/$\eta_{\mathrm{out}}$.  Both exhibit a very strong dependence on $\eta_{\mathrm{in}}$/$\eta_{\mathrm{out}}$ at the boundary to the stable regime.}}
\end{figure}

\textbf{Features of proportionate growth in toe regime.} 
Remarkably, in regime II once a first generation of toes has developed, the instability is again suppressed; no further splitting is observed, even though the toe width has grown much larger than $2\lambda_{\mathrm{c}}$ $\sim$ 10$b$, where one would expect tip splitting.  The regime over which the interface becomes unstable only once and then grows stably depends on the flow rate $q$; as $q$ increases new generations of toes start to appear at lower viscosity ratios within the duration of the experiment. For typical flow rates used in the experiments, this regime comprises the range 0.1 $<$ $\eta_{\mathrm{in}}$/$\eta_{\mathrm{out}}$ $<$ 0.3. We find that within this range, the pattern evolution exhibits the unusual features characteristic of proportionate growth.  

The most common example of proportionate growth is the growth of mammals; as a baby mammal grows, different parts of its body grow at the nearly the same rate and thus in direct proportion to each other.  The head also grows but at a somewhat slower rate.  This keeps their overall shapes unchanged.  How the body organizes this synchronized growth is a longstanding question for biologists. The key features of proportionate growth are that the growing patterns are composed of distinguishable structures with sharp boundaries, all of which grow at nearly the same rate, keeping their overall shapes unchanged. A certain robustness to external noise is further required. Sadhu and Dhar have recently shown that in an abelian sandpile model proportionate growth comes out naturally from local rules, and they propose that proportionate growth may be an example of self-organization\cite{Sadhu2012,Dhar2013}.  However, examples of proportionate growth in real physical systems outside the biological world are difficult to find.  

The patterns formed in the toe regime provide a rare physical (as distinct from biological) example of such a growth process. We recall that while the toes initially grow faster than the inner radius $R_{\mathrm{i}}$, their growth slows down at later times and becomes almost proportional to the growth of the inner radius; the ratio of the two distinct length scales becomes only very weakly dependent on time. In addition, in the proportionate growth regime described above, only one generation of toes develops so that the first generation of structure does not further split into new generations. To demonstrate that the patterns are indeed very similar during the growth process, Fig. 3a shows snapshots of the pattern at three times (top row) and magnifications of those same patterns so that they all have the same outer radius in the image (bottom row). The enlarged images look essentially indistinguishable. The characteristics of proportionate growth observed in two dimensions are further reflected in the third dimension. This is shown in Fig. 3b, where we show the cross-sectional profiles of the toes reported in Fig. 3a. The shape of the toes across the gap remains self-similar during the growth process, as evidenced by the reasonable scaling of the three profiles upon normalisation with the total size of the toe, $L_{\mathrm{c}}$. The range of sizes over which we observe proportionate growth in our system corresponds to a growth of the toes by about a factor of four within the duration of an experiment. This exceeds the factor by which a human typically grows. We note that our patterns are robust to a certain level of noise; neither small air bubbles nor tiny filaments deliberately placed on the glass plates influence the patterns.

\begin{figure}
\centering
\includegraphics[scale=0.5]{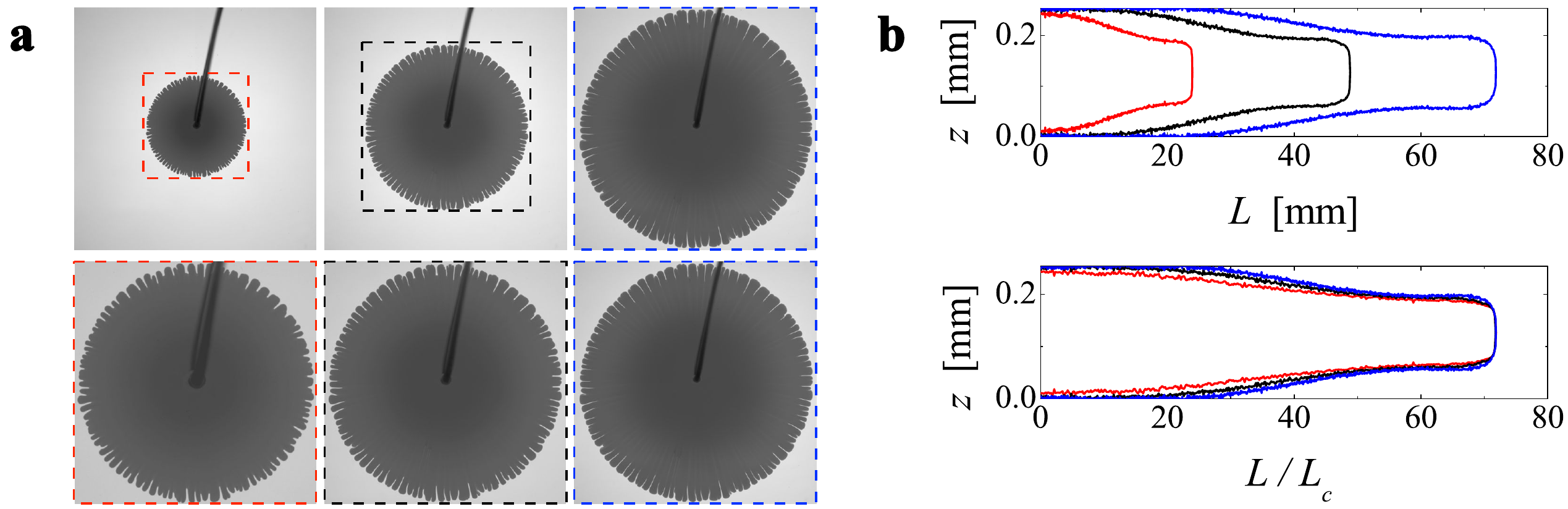}
\caption{ \textnormal{\textbf{Features of proportionate growth in regime II.} \textbf{a},  Top row: Temporal evolution of a miscible fingering pattern with $\eta_{\mathrm{in}}$/$\eta_{\mathrm{out}}$ = 0.185. The scale bar is 5 cm. Bottom row: Zoomed images of the fingering pattern, where the images are enlarged to have the same outer radius. The enlarged region is indicated by the coloured squares. The zoomed images are essentially indistinguishable, indicating that the toes grow in direct proportion to the overall pattern. \textbf{b}, Cross-sectional profiles of the patterns shown in \textbf{a}. The shape of the finger across the gap remains self-similar during the growth process (top), as evidenced by the fair superposition of the profiles when normalised with the total size of the toe $L_{\mathrm{c}}$ (bottom). The growth occurs proportionate in all three dimensions.}}
\end{figure}

\clearpage 

\textbf{Discussion}\\
The remarkably large range of distinct patterns observed in the small interfacial-tension limit reveals a novel aspect of the Saffman-Taylor instability beyond the well-studied fingering onset. In particular, it demonstrates that the efficiency of outer fluid displacement is drastically changed by using fluids with different viscosity ratios as distinct from different viscosity differences. Strikingly, the viscosity ratio does controls not merely the rate of pattern growth, but also its very nature. Indeed, by tuning this single parameter, the system transitions from fractal growth with characteristics of DLA, where growth occurs only at the edge of the pattern, to proportionate growth, where multiple lengthscales all grow at the same rate preserving the overall shape.

The significant delay in the onset of the instability with increasing viscosity ratio is surprising. Indeed, classical stability analysis arguments developed extensively over the past few decades predict in the absence of surface tension an onset radius of the instability that is much smaller than that observed in our experiments. An evaluation of when the maximal growth rate first becomes positive leads to a viscosity-ratio-dependent onset radius between 1$b$ and 4$b$ in the viscosity range of interest (0.166 $<$ $\eta_{\mathrm{in}}$/$\eta_{\mathrm{out}}$ $<$ 0.5)\cite{Paterson1981,Nagel2013}. This value is smaller than the size of the central injection hole in our plates and would correspond to timescales that we could not resolve in our experiments. 

Both $t_{\mathrm{c}}$ and ($R_{\mathrm{f}}$/$R_{\mathrm{i}}$)$_{\mathrm{c}}$ exhibit a dramatic variation approaching the boundary between the stable and unstable regimes, which suggests a complete elimination of the instability at high viscosity ratios. Further, the boundary coincides with a change in the 3-D shape of the ``tongues'' from rounded to flat fronts, as mentioned previously.  Lajeunesse et al.\cite{Lajeunesse1997,Lajeunesse1999} have shown that in the limit of long times and zero diffusion, the transition between a smooth profile and a shock front is analytically predicted to occur at $\left.\eta_{\mathrm{in}}/\eta_{\mathrm{out}}\right|_{\mathrm{c}}$ = 0.67. This value is higher than that observed in our experiments: $\left.\eta_{\mathrm{in}}/\eta_{\mathrm{out}}\right|_{\mathrm{c}}$ $\approx$ 0.33. The experiments however cannot probe the long-time limit. The non-linear growth of the patterns we have analysed can only occur once the instability is triggered. It is important to point out that both the connection between the shock-front formation and the onset of the lateral instability and the suppression of any instability (\emph{e.g.}, of the kind from the original  Saffman-Taylor analysis) for 0.67 $<$ $\eta_{\mathrm{in}}$/$\eta_{\mathrm{out}}$ $<$ 1 remain to be explained.

That the viscosity ratio is an important control parameter for determining $R_{\mathrm{f}}$/$R_{\mathrm{i}}$ can be seen in a simple analysis of pressure drops after the instability has been fully established (see Methods). There, we consider two neighbouring channels each filled with fluids having the same inner and outer viscosities. The interface between the two fluids is located at different positions along each channel. At some distance ahead of the leading interface, the pressure drop is equalised in the two channels. In that case, the interfaces will move at different velocities that depend on $\eta_{\mathrm{in}}$/$\eta_{\mathrm{out}}$. Since the distance at which pressure equalisation occurs can be different in linear or circular geometries, new experiments in a linear cell should be performed to see which features we have found in the radial geometry are robust.

An important conclusion can be drawn from this analysis. The mechanism by which the viscosity ratio controls the growth of $R_{\mathrm{f}}$/$R_{\mathrm{i}}$ does not require any three-dimensional effects. This implies that the effect of the viscosity ratio in setting the large-scale pattern is distinct from the effect of the viscosity ratio on the shape of the tongue across the gap. This is confirmed by experiments showing that the viscosity ratio similarly sets $R_{\mathrm{f}}$/$R_{\mathrm{i}}$ for \emph{immiscible} fluids, which do not display such complex structure in the third dimension (manuscript in preparation). 

In conclusion, our experiments probe the viscous fingering instability using miscible liquids. In this zero interfacial tension limit, the most-unstable wavelength is minimal, which would suggest highly unstable behaviour. However, the experiments show a different outcome with the formation of different types of large-scale structures. This large-scale aspect of the patterns is governed by the viscosity ratio between the two liquids and is independent of the most-unstable wavelength that determines the onset of the instability locally at the interface. In particular, at high viscosity ratios a regime of completely stable displacement exists\cite{Lajeunesse1997, Lajeunesse1999} showing that the removal of the stabilizing interfacial tension in fact stabilizes the interface.

Our studies reveal a much richer pattern formation in the small-wavelength limit than had previously been established in experiments performed at low viscosity ratios and large pressure gradients, where the patterns are characterized by highly branched fingers with the same fractal dimension $d_{\mathrm{f}}$ = 1.7 as found in diffusion-limited aggregation\cite{Witten1981,Nittmann1985,Daccord1986,Mathiesen2006,Praud2005,Cheng2008}.  In particular, with increasing viscosity ratio the length of the fingers drastically decreases and effectively produces a much more efficient displacement of one fluid by the other. This diversity of possible patterns can lead to new routes to controlled fluid displacements with the viscosity ratio between the fluids as a convenient control parameter.

\begin{methods}
\textbf{Experiments.} 
The miscible fluids used here are mixtures either of water and glycerol or of two mineral oils (Fisher Scientfic). Water and glycerol are miscible in all proportions, allowing access to viscosities between 1 mPa s and 1350 mPa s. The viscosity of water-glycerol mixtures changes non-linearly with concentration: at high glycerol concentrations, a large viscosity change is achieved with very small variation in chemical composition, substrate wetting or density\cite{Lide}. Because of the high viscosities, the inter-diffusion of the fluids is negligible; thus during an experiment the fluids remain separated by a well-defined interface\cite{Lajeunesse1999,Yang1997,DErrico2004}. Mineral oil and water-glycerol mixtures produce identical results indicating that substrate wetting and chemical interactions between the liquids do not affect pattern formation. 

Our experiments are performed in a Hele-Shaw geometry consisting of two 1.9 cm thick circular glass plates of diameter $L$ = 28 cm. The gap between the plates, $b$, can be varied between 76 $\mu$m and 1143 $\mu$m, and is maintained uniform in an experiment to within 1-2\% by spacers around the perimeter. The liquids are pumped through a 1.6 mm hole in the centre of the top plate with a syringe pump (SyringePump NE-1010) at constant volumetric flow rates between 0.4 ml min$^{-1}$ and 40 ml min$^{-1}$. The patterns are recorded with a Prosilica GX 3300 camera at frame rates up to 15 frames per sec.

To measure the profile across the gap, we convert the measured optical absorption of the dyed inner liquid (brilliant blue G, Alfa Aesar) to liquid thickness. For the unstable patterns, the cross-section is taken along the middle of a finger. The tongues formed by the inner fluid are confined to the gap centre. This is confirmed by experiments where we use a UV-curable polymeric fluid as one of our fluids. This allows us to ``freeze'' the 3D patterns by exposure to UV light.

\textbf{Analysis of pressure drop.} 
Our experiments investigating the large-scale patterns that form upon invasion of one fluid into another of higher viscosity reveal that the viscosity ratio $\eta_{\mathrm{in}}$/$\eta_{\mathrm{out}}$ is a control parameter for the global features of the instability patterns. In particular, we show that $\eta_{\mathrm{in}}$/$\eta_{\mathrm{out}}$ sets the ratio of the finger size to the size of the inner radius, $R_{\mathrm{f}}$/$R_{\mathrm{i}}$. 

A simple analysis of the pressure drop in the system after the instability has been fully established captures the important role of $\eta_{\mathrm{in}}$/$\eta_{\mathrm{out}}$ in governing the pattern growth. We here consider two neighbouring channels each filled with fluids of viscosity $\eta_{\mathrm{in}}$ and $\eta_{\mathrm{out}}$, as shown in Fig. 4a and b for a linear geometry and a radial geometry, respectively. The interface between the two fluids is located at $R_{\mathrm{o}}$ in channel 'o' and at $R_{\mathrm{i}}$ in channel 'i'. We assume that the pressure is equalised in the two channels at a certain distance behind $R_{\mathrm{i}}$ at position $r_{\mathrm{p}}$ and at a certain distance ahead of $R_{\mathrm{o}}$ at position $r_{\mathrm{p}-\Delta \mathrm{p}}$. 

\begin{figure}[htb]
\centering
\includegraphics[scale=0.7]{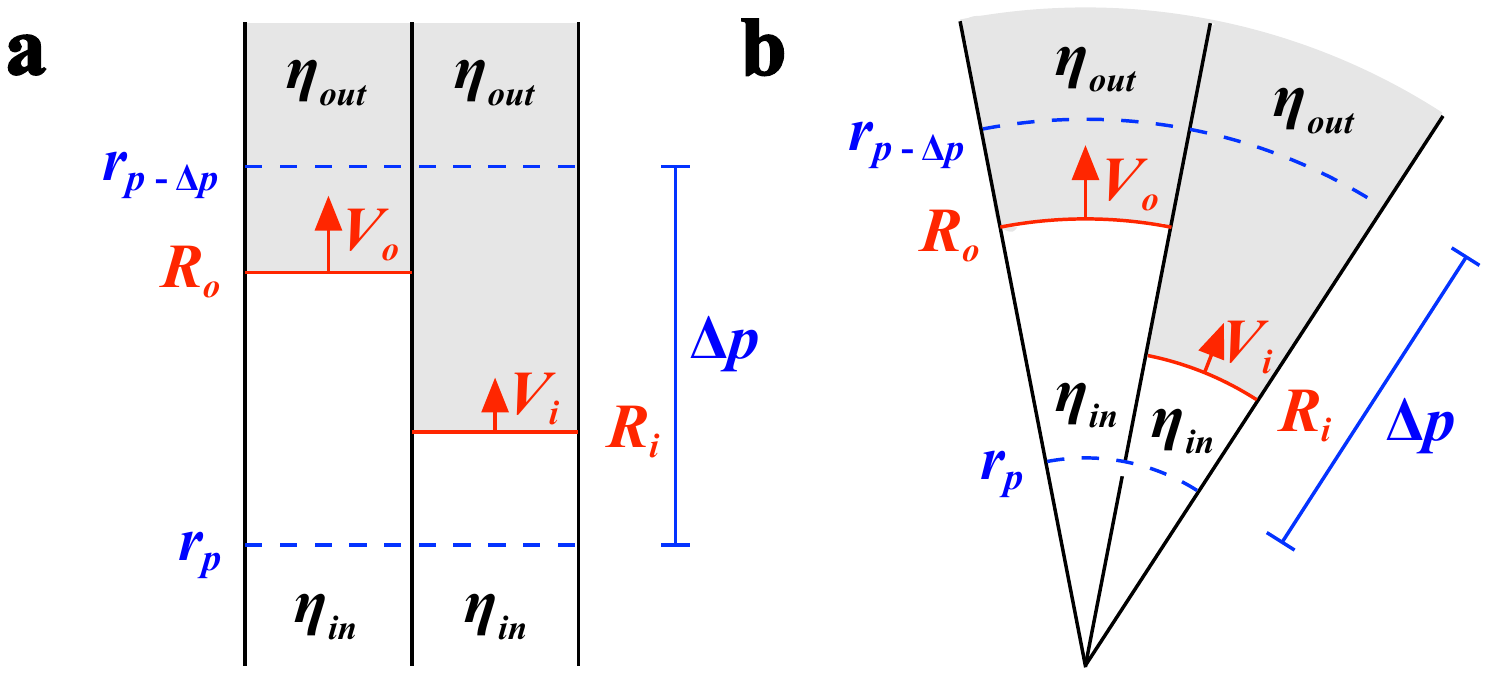}
\caption{ \textnormal{\textbf{ Schematic representation of the channel geometry.} Two channels, of either \textbf{a}, linear geometry or \textbf{b}, radial geometry are each filled with an outer fluid of viscosity $\eta_{\mathrm{out}}$ which is displaced by an inner fluid with $\eta_{\mathrm{in}}$. The interface between the two fluids is located at $R_{\mathrm{\mathrm{o}}}$ in channel 'o' and at $R_{\mathrm{i}}$ in channel 'i'. The two channels experience an identical pressure drop of $\Delta$$p$ over the distance from $r_{\mathrm{p}}$ to $r_{\mathrm{p}-\Delta \mathrm{p}}$. }}
\end{figure}

For the linear geometry, we can calculate the pressure drop $\Delta p$ from $r_{\mathrm{p}}$ to $r_{\mathrm{p}-\Delta \mathrm{p}}$ in each channel using Darcy's law\cite{Batchelor2000}. 
\begin{equation}
\Delta p_{\mathrm{o}}(t) \propto \eta_{\mathrm{in}} V_{\mathrm{o}}(t) \big[R_{\mathrm{o}}(t)-r_\mathrm{p}(t)\big] + \eta_{\mathrm{out}} V_{\mathrm{o}}(t) \big[r_{(\mathrm{p}-\Delta \mathrm{p})}(t)-R_{\mathrm{o}}(t)\big]
\label{eq:Cross}
\end{equation}
\begin{equation}
\Delta p_{\mathrm{i}}(t) \propto \eta_{\mathrm{in}} V_{\mathrm{i}}(t) \big[R_{\mathrm{i}}(t)-r_\mathrm{p}(t)\big] + \eta_{\mathrm{out}} V_{\mathrm{i}}(t) \big[r_{(\mathrm{p}-\Delta \mathrm{p})}(t)-R_{\mathrm{\mathrm{i}}}(t)\big]
\label{eq:Cross}
\end{equation}
where $V_{\mathrm{o}}$$(t)$ and $V_{\mathrm{i}}$$(t)$ are the velocities of the interface in channel 'o' and 'i', respectively.
The pressure drop is the same in both channels: 
\begin{equation}
\Delta p_{\mathrm{o}}(t) = \Delta p_{\mathrm{i}}(t) 
\label{eq:Cross}
\end{equation}
Therefore:
\begin{equation}
\eta_{\mathrm{in}} V_{\mathrm{o}}(t) \big[R_{\mathrm{o}}(t)-r_\mathrm{p}(t)\big] +  \eta_{\mathrm{out}} V_{\mathrm{o}}(t) \big[r_{(\mathrm{p}-\Delta \mathrm{p})}(t)-R_{\mathrm{o}}(t)\big] = \eta_{\mathrm{in}} V_{\mathrm{i}}(t) \big[R_{\mathrm{i}}(t)-r_\mathrm{p}(t)\big] + \eta_{\mathrm{out}} V_{\mathrm{i}}(t) \big[r_{(\mathrm{p}-\Delta \mathrm{p})}(t)-R_{\mathrm{i}}(t)\big]
\label{eq:Cross}
\end{equation}
\begin{equation}
V_{\mathrm{o}}(t) \Big[(\frac{\eta_{\mathrm{in}}}{\eta_{\mathrm{out}}}-1) R_{\mathrm{o}}(t) -\frac{\eta_{\mathrm{in}}}{\eta_{\mathrm{out}}}r_\mathrm{p}(t) +  r_{(\mathrm{p}-\Delta \mathrm{p})}(t)\Big] = V_{\mathrm{i}}(t)  \Big[(\frac{\eta_{\mathrm{in}}}{\eta_{out}}-1) R_{\mathrm{i}}(t) -\frac{\eta_{\mathrm{in}}}{\eta_{out}}r_\mathrm{p}(t) +  r_{(\mathrm{p}-\Delta \mathrm{p})}(t) \Big]
\label{eq:Cross}
\end{equation}

This shows that $\eta_{\mathrm{in}}$/$\eta_{\mathrm{out}}$ sets the velocity of the interface in each channel and therefore the growth of $R_{\mathrm{o}}$ and $R_{\mathrm{i}}$. Clearly, this description is oversimplified but it does capture the essential features of how pressure affects the velocity in the two channels and it directly reveals the importance of $\eta_{\mathrm{in}}$/$\eta_{\mathrm{out}}$ in governing the large-scale patterns. From a similar analysis the same conclusion can be drawn for the radial geometry. However, depending on the geometry used the distance over which the pressure equalises in the two channels can be different. This highlights the need for experiments performed in a linear cell to investigate the robustness of the features observed in our experiments performed in radial cells.

\end{methods}

\clearpage 


\begin{addendum}
 \item We thank Rudro Rana Biswas, Justin Burton, Deepak Dhar, Todd Dupont, Julian Freed-Brown, Leo Kadanoff, Paul Wiegmann, Tom Witten and Wendy Zhang.  This work was supported by NSF Grant DMR-1404841.  I. B. gratefully acknowledges financial support from the Swiss National Science Foundation (PBFRP2-134287).
 \item[Author Contributions] I.B., R.R. and S.R.N. designed research; I.B. and R.R. performed research;  I.B., R.R. and S.R.N. analysed data; I.B. and S.R.N. wrote the paper.
 \item[Competing Interests] The authors declare no competing financial interests.
 \item[Correspondence] Correspondence should be addressed to Irmgard Bischofberger. \\*~(email: ibischofberger@uchicago.edu).
\end{addendum}

\end{document}